\pdfoutput=1
\documentclass[12pt]{iopart}

\usepackage{iopams}

\usepackage{epstopdf}
\usepackage{psfrag}
\usepackage{ps4pdf}

\usepackage{multirow}
\usepackage{paralist}
\usepackage{color}

\begin{document}

\newcommand{\D}{\mathrm{d}}
\newcommand{\pr}{{}^{\tiny{\mbox{ïð}}}}
\newcommand{\nuint}{\nu_{\mbox{\footnotesize{èíò}}}}
\newcommand{\Vreg}{V_{\mbox{\footnotesize{ðåã}}}}

\newcommand{\OB}[2]
{\overbrace{\hspace{#1mm}}^{ #2}}

\newcommand{\zd}
{
\scriptstyle{b-z(\tau)}\;
 \Bigg\{
}

\newcommand{\BC}{
\OB{22}{\frac{b}{c}}
}

\newcommand{\ZC}{
\OB{10}{\frac{b-z(\tau)}{c}}
}

\title{Relativity, Doppler shifts, and retarded times
 in deriving the correction for the finite speed of light: a comment on 'Second-order Doppler-shift corrections in free-fall absolute gravimeters' }
\author{V D Nagornyi}
\address{Metromatix, Inc., 111B Baden Pl, Staten Island, NY 10306, USA}
\ead{vn2@member.ams.org}
\author{Y M Zanimonskiy}
\address{Institute of Radio Astronomy, National Academy of Sciences of Ukraine, 4, Chervonopraporna St., Kharkiv, 61002, Ukraine.}
\author{Y Y Zanimonskiy}
\address{International Slavonic University, 9-A, Otakara Jarosha St., Kharkiv, 61045, Ukraine.}
\begin{abstract}
In the article (\emph{Rothleitner and Francis} 2011 \emph{Metrologia} \textbf{48} 187-195)
%\cite{rothleitner2011}
the correction due to the finite speed of light in absolute gravimeters is analyzed from the viewpoint of special relativity. The relativistic concepts eventually lead to the two classical approaches to the problem: analysis of the beat frequency, and introduction of the retarded times. In the first approach, an additional time delay has to be assumed, because the frequency of the beam bounced from the accelerated reflector differs at the point of reflection from that at the point of interference. The retarded times formalism is equivalent to a single Doppler shift, but results in the same correction as the beat frequency approach, even though the latter is explicitly combines two Doppler shifts. In our comments we discuss these and other problems we found with the suggested treatment of the correction.
\end{abstract}
%
%
%\tableofcontents
%
%
\newpage
\pagestyle{empty}
\section{Introduction}
Correction due to the finite speed of light is one of the most controversial subjects in absolute gravimetry. For a long time the inconsistencies in the results published by different authors did not get proper attention. It's therefore very satisfying that shortly following our paper on this subject \cite{nagornyi2011}, another group of authors has also published their view on the problem \cite{rothleitner2011}. In our paper, we uncovered the reasons of discrepancies in the results obtained over the years, and found that some currently used formulas incompletely account for the initial velocity of the test mass, leading to the error of up to 1 $\mu$Gal. The findings of our counterparts, however, are a way more significant. According to the authors, modern gravimeters overcorrect the influence of the finite speed of light by about 4 $\mu$Gal \cite{rothleitner2011}. Discovering a systematic error that big would be very important, as some intercomparisons of absolute gravimeters, especially those with different instrument types involved, show significant discrepancy in the results \cite{louchetchauvet2010}.
%In analysis of the problem as contradictory as the finite speed of light correction, we can only applaud to the approach the authors of \cite{rothleitner2011} have taken by building their reasoning up from the ground, starting with the wave equation, then going to the two-beam interferometry, and so on up to the disturbances of the test body's trajectory.
However, upon scrutinizing the analysis done in \cite{rothleitner2011}, we had to admit that the suggested reasoning not only implements
%(thought, in more sophisticated way)
the same misconceptions that flawed the conclusions of other authors, but also introduces some new ones. We now explain our opinion by considering some old and new misconceptions in more details.
\section{Relativity is irrelevant for the analysis of  corrections in absolute gravimeters}
The Taylor expansion of the Lorenz factor
\begin{equation}
\label{Lorendz}
\gamma = \frac{1}{\sqrt{1-\frac{V^2}{c^2}}}
= 1+\frac{V^2}{2c^2} + O\left(\frac{V^4}{c^4} \right)
\end{equation}
allows to estimate the effects of special relativity, if ever applicable, to absolute gravimeters. The maximum velocity $V$ of the test mass in modern instruments does not exceed 2 m/s, so the component $\frac{V^2}{2c^2}$ is only about $10^{-17}$. This value is not only much lower than the practically reachable accuracy of $10^{-10}$ \cite{faller2002}, but is also below the Heisenberg's uncertainty as applied to ballistic gravimeters \cite{bondarev1986eng}. The principles of special relativity are therefore redundant for the analysis done in \cite{rothleitner2011}. The facts the authors use to analyze the correction, such as the double Doppler shift (formula (9)), or the time delay (formula (35)) are direct results of the classical mechanics.
\section{For the accelerated motion, additional time delay is necessary besides the double Doppler shift to correctly account for the finite speed of light}
\label{sec_extra_delay}
The double Doppler shift formula
\begin{equation}
\label{f_reflected_V}
f_0'' = f_0 \frac{1+\frac{V}{c}}{1-\frac{V}{c}}
\end{equation}
relates the frequency $f_0$ of the laser beam directed towards the approaching with the velocity $V$ reflector, to the frequency $f_0''$ of the reflected beam. Implementation of this formula to find the beat frequency $(f_0 - f_0'')$ is significantly different for uniform or accelerated motion of the reflector.

If the reflector moves with the constant velocity $V_0$, the frequencies $f_0$ and $f_0''$ do not change in time (fig.\ref{DOPPLER_EFFECTS}a),
\begin{figure}[h]
\centering
%\singlespacing
\small
\input{DOPPLER_EFFECTS.pic}
  \caption[short title]
  {
  \quad\parbox[t]{11cm} {Frequency of the reflected laser beam for different modes of the reflector's motion:
      \begin{compactdesc}
      \item[a]{--- uniform motion --- the reflected frequency does not change in time;}
      \item[b]{--- accelerated motion --- the reflected frequency is higher at the point of reflection due to the finite speed of the light wave propagation.}
      \end{compactdesc}
  }
  }
\label{DOPPLER_EFFECTS}
\end{figure}
therefore the beat frequency, as the authors of \cite{rothleitner2011} point out in the formula (10), for any moment of time is defined by
\begin{equation}
\label{RF_10}
\Delta f_0'' = \left[ \frac{2V_0}{c}+\frac{2V_0^2}{c^2} + O\left(\frac{V_0^3}{c^3} \right) \right] f_0
.
\end{equation}
In case of the accelerated motion, the reflected frequency does change in time.
%, but the notation $f_0''$ does not specify which moment of time the value is considered at.
The frequency corresponding to the double Doppler shift %(\ref{f_reflected_V})
occurs at the moment of the reflection $\tau$, while the interference occurs at the later moment $t$ (fig.\ref{DOPPLER_EFFECTS}b), and \emph{the frequency is different at the moments $\tau$ and $t$}. Ignoring this fact would imply that the increased frequency propagates instantaneously, that is, the speed of light is infinite. So, to get the beat frequency for the accelerated motion of the reflector, it's not enough just to replace $V_0$ with $V_0 + g_0t$ at (\ref{f_reflected_V}), we also need to assign the appropriate timing to the resulting process.
%moments: either  $t$ or $\tau$ to every term.
The author's formula (14) for the frequency at the moment of reflection should actually be written as
\begin{equation}
\label{RF_14_modified}
f_0''(\tau) = \left(1+
\frac{2V_0}{c} + \frac{2g_0}{c} \tau + \frac{2V_0^2}{c^2} + \frac{4V_0}{c^2}g_0 \tau + \frac{2g_0^2}{c^2} \tau^2
\right) f_0,
\end{equation}
so that the beat frequency $(f_0''(\tau)-f_0)$ is obtained \emph{as if} the interference took place at the moment of reflection. This beat frequency, mistakenly used for the analysis in \cite{rothleitner2011}, corresponds to the following acceleration:
\begin{equation}
\label{RF_14_modified_acc}
g(\tau) = \frac{\partial}{\partial \tau} \left(\frac{\lambda}{2}(f_0''(\tau) - f_0)\right)
= g_0 + \frac{2g_0}{c}(V_0 + g_0 \tau)
,
\end{equation}
where $\lambda$ is the laser wavelength.
Because the interference is actually happens at the beam splitter, the frequency should be taken at the moment $t$, not $\tau$. Expressing $\tau$ via $t$ with
\begin{equation}
\label{t_tau}
\tau =t-\frac{b-V_0-g_0 t^2/2}{c}
\end{equation}
and substituting it to the right side of (\ref{RF_14_modified}), we find the reflected frequency $f_0''(t) $ at the moment of the interference. The acceleration found similar to (\ref{RF_14_modified_acc}) will then be
\begin{equation}
\label{cor_acc}
g(t) = g_0 + \frac{3 g_0}{c}(V_0 + g_0 t).
\end{equation}
The above formula represents the disturbed acceleration that should be used to derive the correction for the finite speed of light.
\section{The gravimeter data model is essential for deriving corrections}
\begin{figure}[h]
\centering
%\singlespacing
\small
\input{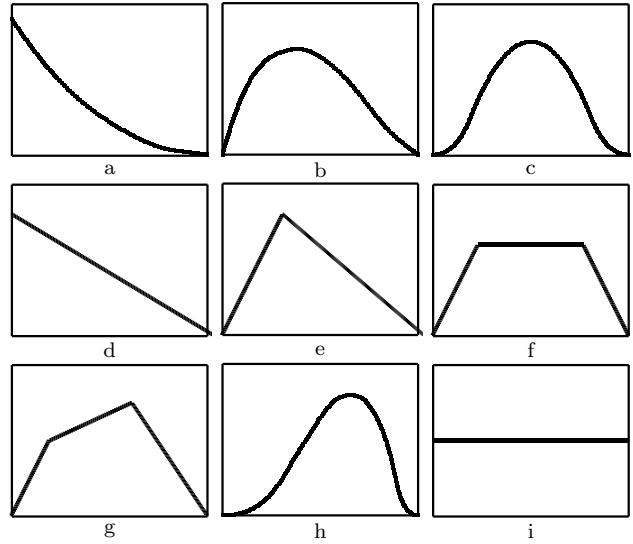}
  \caption[short title]
  {
  \quad\parbox[t]{11cm} {Weighting functions of absolute gravimeters
      \begin{compactdesc}
      \item[a ]{ -- model $S_i = g_0 T_i^2/2$, multiple levels equally spaced in time (EST)}
      \item[b ]{ -- model $S_i = V_0 T_i + g_0 T_i^2/2$, EST}
      \item[c ]{ -- model $S_i = S_0+V_0 T_i + g_0 T_i^2/2$, EST}
      \item[d ]{ -- 2-level schema}
      \item[e ]{ -- 3-level schema}
      \item[f ]{ -- 4-level schema}
      \item[g ]{ -- the model $S_i = S_0+V_0 T_i + g_0 T_i^2/2$ fit to 4 levels}
      \item[h ]{ -- the model $S_i = S_0+V_0 T_i + g_0 T_i^2/2$, multiple levels equally spaced in distance (ESD)}
      \item[i ]{ -- the uniform weighting function once believed to belong to the above model (\textbf{h}) \cite{zumberge1981}}
      \end{compactdesc}
  }
  }
\label{WEIGHTING_FUNCTIONS}
\end{figure}

The expression (\ref{cor_acc}) defines the observable acceleration, disturbed by the effects of the finite speed of light. If the measurand is calculated in the gravimeter linearly with respect to the distance intervals, then any disturbed acceleration
\begin{equation}
\label{acc_disturbed}
g(t) = g_0 + \Delta g(t)
\end{equation}
translates to the measured acceleration according to the formula \cite{nagornyi1995}
\begin{equation}
\label{acc_measured}
\overline g = \int_0^T \left(g_0 + \Delta g(t)\right)w(t) \D t =
g_0 + \int_0^T \Delta g(t)w(t) \D t,
\end{equation}
where $\Delta g(t)$ is a disturbance acting during the measurement interval $T$, $w(t)$ is the weighting function of the gravimeter. The correction for the disturbance $\Delta g(t)$ is found as
\begin{equation}
\label{acc_correction}
\overline {\Delta g} = -\int_0^T \Delta g(t)w(t) \D t.
\end{equation}
The weighting function depends on the gravimeter's data model that includes
\begin{itemize}
\item the location of the time-distance pairs along the trajectory (measurement schema);
\item the formula used to calculate the acceleration using the time-distance pairs (working formula).
\end{itemize}
Some weighting functions for different data models are shown on the fig.\ref{WEIGHTING_FUNCTIONS}.
The weighting function always obeys the unit square property:
\begin{equation}
\label{unit_square}
\int_0^T w(t) \D t \equiv 1,
\end{equation}
which ensures that the component $g_0$ of (\ref{acc_measured}) is estimated bias-free for any linear working formula. However, the diversity of weighting functions creates a strong model dependency for the correction (\ref{acc_correction}). That's why any inference of corrections can not be agnostic to the gravimeter's data model. This means that the model used to calculate $g_0$ should agree with the model used to derive corrections. Unless the agreement is observed, the corrections would not adequately compensate the disturbances for which they have been derived. An example of such inter-model disagreement can be found in the paper \cite{murata1978}. The gravity value is calculated by the least squares fit of the model $S_i=z_0+V_0 T_i + g_0 T_i^2/2$ using multiple equally spaced in time (EST) levels (weighting function on fig.\ref{WEIGHTING_FUNCTIONS}c). For the gradient correction, the continuous least squares fit of the simplified model $S_i=g_0 T_i^2/2$ was used\footnote{formula (7-19) of \cite{murata1978}} which corresponds to the weighting function on fig.\ref{WEIGHTING_FUNCTIONS}a. The finite speed of light correction was based on the 2-level model\footnote{formula (7-11) of \cite{murata1978}} with the weighting function shown on the fig.\ref{WEIGHTING_FUNCTIONS}d. As the result, both corrections introduce a bias of several $\mu$Gals.
%
% Another example of the model disagreement can be found in \cite{zumberge1981}, where the gravity is calculated as in previous paper, but the data are equally spaced in distance (ESD) (weighting function on \ref{WEIGHTING_FUNCTIONS}h), but all the corrections were obtained for the  uniform weighting function (\ref{WEIGHTING_FUNCTIONS}i).

As the corrections can not be obtained without the data models, so the results and approaches of different authors can not be compared without specifying the models they used. In some cases, special steps have to be taken to achieve the model agreement and enable the comparison \cite{svitlov2010a}.
In \cite{rothleitner2011}, however, both the corrections are derived
%(no resulting formula for the correction is given, though)
, and the results are compared with no model involvement. Some model information was probably supposed to be incorporated in the new ``least squares correction'' introduced in the paper. As the term suggests, the models not using the least squares fit\footnote{For example, the 2-level model used in the paper \cite{murata1978} to derive the correction} are left aside. There are, however, bigger problems with this correction, which we discuss in the following section.
\section{Trajectory disturbance: a number or a process? and the ``least square correction''}
\label{sec_number_or_process}
The disturbance $\Delta g(t)$ of the observed free falling mass motion (\ref{acc_disturbed}) is a function, while the correction (\ref{acc_correction}) is a number. Not necessarily the disturbance has to be expressed in terms of acceleration --- it can as well be expressed as changing coordinate, or velocity, but it's always a process in time. Analyzing the speed of light correction, the authors of \cite{rothleitner2011} do not make clear distinction between the disturbance and the correction.  In the formula (24), for example, the disturbed coordinate is called the ``trajectory bias'' and found as definite integral of the beat frequency over the entire measurement interval, i.e. as number. On the other hand, in the formula (25) the authors find the second derivative of the bias in $T$, thus treating it as function. Most avidly the ''process vs number'' confusion is revealed in the formulas (60)--(62), where the left sides are functions of time $\Delta z(t)$, but the time argument $t$ is missing on the right side. The missing $t$ is not a typo, as the expressions are obtained as definite integrals in $t$ over the measurement interval $T$.

The ``least squares correction'' introduced for the bias does not resolve its  process/number dualism. First, the authors give an example related to the vertical gravity gradient, saying that if a model without the gradient is fitted to the data containing the gradient, the fitted value of $g_0$ will refer to a certain point between the start and the end of the drop. Then the authors generalize the example saying that \emph{``this kind of correction can be done for any perturbation.''} Such heuristic definition of the correction by example and analogy explains neither \emph{what} is corrected, nor \emph{why}, nor \emph{how}.

Based on the efforts described in the section 5 of \cite{rothleitner2011}, we can suggest that actual idea behind the ``least square correction'' was to determine how the disturbed acceleration translates into the measurement bias in case the least squares fit is applied to calculate the result. Using the weighting functions, this approach can indeed be very efficient, but requires expressing a disturbance as polynomial process in time, and then replacing its coefficients with corresponding averaging coefficients \cite{nagornyi1995}.
This technique, however, does not create any new kind of correction, but just helps to evaluate how the known disturbances translate into corrections.
\section{The retarded times, while equivalent to a single, not double Doppler shift, still lead to the same correction}
\begin{figure}[h]
\centering
%\singlespacing
\small
\input{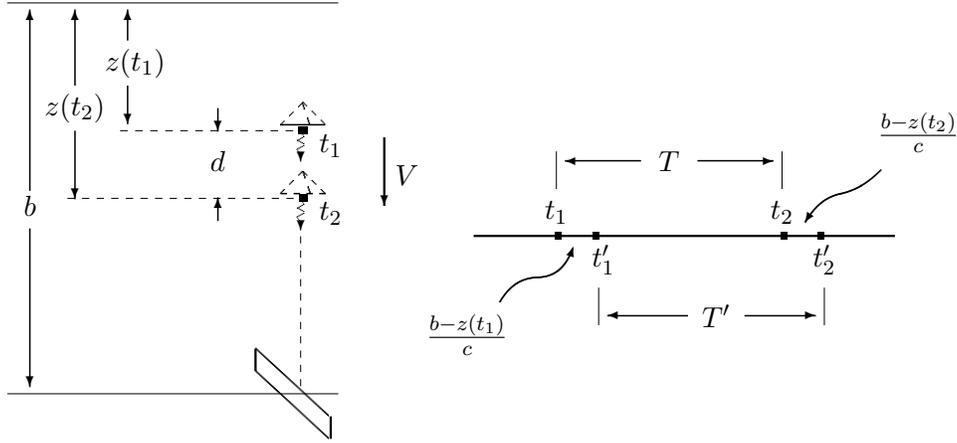}
%
%%%%%%%%%%%%%%%%%%%%%%%%%%%%%%%%%%%%%%%%%%%
%
%   ÏÐÈ ÊÎÌÏÈËßÖÈÈ ÓÄÀËÈÒÜ ÈÇ ÔÀÉËÀ ÑÒÐÎÊÈ:
%
%
%\qbezier(41,38.86)(41,38.86)(41,38.86)
%\qbezier(41,38.86)(41,38.86)(41,38.86)
%
%...
%
%\qbezier(41,29.86)(41,29.86)(41,29.86)
%\qbezier(41,29.86)(41,29.86)(41,29.86)
%
%
%%%%%%%%%%%%%%%%%%%%%%%%%%%%%%%%%%%%%%%%%%%%
%
  \caption[short title]
  {
  \quad\parbox[t]{11cm} {Distortion of the measured time intervals due to the finite speed of light
  }
  }
\label{DELAYS}
\end{figure}

Let's consider the delays experienced by the photons reflected from the test mass at the moments $t_1$ and $t_2$ (fig.~\ref{DELAYS}). Because of the finite speed of light, the photons reach the beam splitter at the later moments $t_1'$ and $t_2'$.
Let's find how the time interval $\Delta t' = t_2'-t_1'$ is related to the time interval $\Delta t = t_2-t_1$.
We have
\begin{equation}
t_1' = t_1 + \frac{b-z(t_1)}{c},
\label{eq_t1}
\end{equation}
\begin{equation}
t_2' = t_2 + \frac{b-z(t_2)}{c},
\label{eq_t2}
\end{equation}
so
\begin{equation}
\Delta t' = t_2'  - t_1' = t_2 - t_1 - \frac{z(t_2)-z(t_1)}{c} =
\Delta t - \frac{d}{c},
\label{eq_T}
\end{equation}
where $d$ is the vertical separation between the positions of the reflector at the moments $t_1$ and $t_2$ (fig.~\ref{DELAYS}). For a small $d$, the observed velocity $V'=d/\Delta t'$ of the reflector is related to its true velocity $V=d/\Delta t$ as
\begin{equation}
V' = \frac{d}{\Delta t'}
= \frac{d}{\Delta t -\frac{d}{c}}
= \frac{1}{\frac{\Delta t}{d}-\frac{1}{c}}=
\frac{1}{\frac{1}{V}-\frac{1}{c}}=
\frac{V}{1-\frac{V}{c}}
\approx
V\left(1+\frac{V}{c}
%+ O\left(\frac{V}{c}\right)^2
\right)
.
\label{eq_V}
\end{equation}
The observed  velocity looks like it has undergone the Doppler shift, even though no shift was considered. While not totally coincidental\footnote{Both the Doppler shift and the signal delay are different manifestations of the finiteness of the speed of light.}, the similarity of the formulas does not enable us to compare physical models based on the implemented number of Doppler shifts and signal delays, because the ways they implemented may vary significantly from one model to another.

Deriving the correction through the beat frequency is based on the interference of the direct and the reflected beams, so two Doppler shifts have to be applied to the signal frequency, plus a single time delay applied to the reflector's velocity, as discussed in the section \ref{sec_extra_delay}. The resulting disturbed acceleration is given by the formula (\ref{cor_acc}).

The retarded times approach uses only one time delay, but applied to the reflectors's coordinate.  The equivalence of this method to the first one can be established by comparing the disturbed accelerations. Assuming $S(t)= z_0 + V_0 t + g_0 t^2 / 2$, we get
\begin{equation}
\label{eq_acc_coord_w_delay}
g(t) = \frac{\partial^2}{\partial t^2}
\left(z(t)\left(1 + \frac{\dot z(t)}{c}   \right)\right) \equiv
g_0 + \frac{3 g_0}{c}(V_0 + g_0 t),
\end{equation}
which is equivalent to (\ref{cor_acc}). This proves that the retarded times approach yields the same correction as the beat frequency one.
\section{Miscellaneous notes}
%
%This section discusses less critical, but still noteworthy issues we find in the paper \cite{rothleitner2011}.
%
\begin{description}
\item
\emph{Title of the paper.}
The second-order Doppler shift is a synonym for the relativistic Doppler shift \cite{mungall1971}, which is irrelevant for absolute gravimeters. In \cite{rothleitner2011} the term is actually used for the second order component of the expansion of the regular Doppler shift.
\item
\emph{Disturbance magnitude.}
The formulas (60) -- (62) of \cite{rothleitner2011} disagree in their orders of magnitude. Only the formula (61) has the magnitude of a disturbance, while two other formulas represent entire disturbed trajectory.
%
%\subsection{On the history of multi-level schemas}
%%
%The authors' statement that the paper \cite{niebauer1989} was first to calculate the corrections for the gravimeters using the least squares fit of the trajectory is historically inaccurate. First papers calculating such corrections in the way similar to \cite{niebauer1989} were published decades earlier \cite{cook1965a, thulin1961}.
%%
%
\item
\emph{The correction by Kuroda \& Mio.}
The authors' comments on the way the correction is obtained in the paper \cite{kuroda1991} are not supported by the content of \cite{kuroda1991}. Nowhere in \cite{kuroda1991} we find that the reflected beam frequency is obtained using a single Doppler shift, as the authors of \cite{rothleitner2011} claim. The factor 2 in the formulas (5) and (5') of \cite{kuroda1991} is a legitimate consequence of the correct use of the double Doppler shift.
\end{description}

\section{Conclusions}
\begin{enumerate}
\item For the test mass velocities used in absolute gravimeters, the influence of the relativistic effects on the measured gravity is below the discernable level and are irrelevant for the analysis of the corrections.
\item The frequency of the beam bounced from the accelerated reflector changes in time and distance, causing the frequency at the reflector be different from that at the beam splitter, so an additional time delay is needed to correctly deduce the acceleration disturbance from the beat frequency.
\item The formulas similar to the Doppler shift appear in the retarded time models because both are based on the finiteness of the speed of light. The beat frequency and the retarded time models represent different aspects of the light nature and can't be validated against each other based on the number of Doppler shifts necessary for their implementation. Both models are equivalent in terms of the disturbed acceleration, which is defined by the formula (\ref{cor_acc}) of this publication.
\end{enumerate}
\section*{References}
%
%
%\bibliography{disser}
%\bibliographystyle{ieeetr}
%

\end{document}